# Direct Observation of Molecular Cooperativity Near the Glass Transition

E. Vidal Russell and N. E. Israeloff

*Department of Physics and Center for Interdisciplinary Research on Complex Systems*

*Northeastern University, Boston, MA 02115,USA*

**The increasingly sluggish response of a super-cooled liquid as it nears its glass transition[1] (just try to pour refrigerated honey) is prototypical of glassy dynamics found in proteins, neural networks, and superconductors. The notion that molecules rearrange *cooperatively* has long been postulated[2] to explain diverging relaxation times and broadened (non-exponential) response. Recently, cooperativity was observed and analyzed in colloid glasses[3] and in simulations of binary liquids well above the glass transition[4]. But nano-scale studies of cooperativity at the molecular glass transition are lacking[5]. Key issues to be resolved include the precise form of the cooperativity, its length-scale[6], and whether broadened response is intrinsic to individual cooperative regions, or arises only from heterogeneity[7-9] in an ensemble of such regions. Here we describe direct observations of molecular cooperativity near the glass transition in poly-vinyl-acetate (PVAc), through nanometer-scale probing of dielectric fluctuations. Molecular clusters switched spontaneously between two to four distinct configurations, producing complex random-telegraph-signals (RTS). Analysis of the RTS and their power spectra shows that individual clusters exhibit both transient dynamical heterogeneity and non-exponential kinetics.**

Dielectric relaxation of a typical super-cooled liquid slows with decreasing temperature and becomes non-exponential near the glass transition. The corresponding peak in the imaginary component of the dielectric susceptibility, $\varepsilon''(\omega)$, becomes broader in frequency, $\omega$, than a single relaxation-time Debye peak[10]. In the heterogeneous picture, this broadened primary ($\alpha$) response arises from cooperative regions[2] (clusters) of typical size, 2-4 nm[6], each with Debye-like response, but with characteristic relaxation times, $\tau$, which differ from cluster to cluster due to disorder. See figure 1a. Whether heterogeneity is the sole source of broadening has been controversial. Dynamically selective experiments have found heterogeneity that is either short-lived[7,8], or long-lived[9] compared with the average relaxation time, while other experiments are argued to support intrinsic broadening or homogeneity[11].

If it were practical to probe the susceptibility of an individual cooperative region within a glass former, many of the outstanding questions might be addressed. Even a slightly larger volume containing a small number of such regions may be useful, since in the heterogeneous picture, the responses of individual clusters will be centered on different frequencies, producing distinguishable features in the susceptibility spectrum. These deviations from macroscopic response would be most pronounced in the high-frequency tail of the primary ($\alpha$)−susceptibility peak, where scaling behavior is often observed[10], and where very few clusters should be contributing (Fig. 1b). One potential problem in probing such a small volume, is that fluctuations (noise) should become relatively large. Fortunately, the noise spectrum contains the same information as the susceptibility and can be derived from it using the fluctuation-dissipation theorem, as long as the sample is in equilibrium[12]. Debye-shaped susceptibility features produce Lorentzian shaped features in the noise spectrum (Fig. 1c). If observed, these spectral features would provide direct evidence for heterogeneity and cooperativity and make new details of glassy dynamics accessible.

By sensing local electrostatic forces, non-contact atomic force microscopy (AFM) techniques can be used to measure local dielectric properties[13]. Recently, dielectric relaxation[14] and spontaneous thermal dielectric noise[15] were studied in glassy polymers with non-contact AFM techniques. In the present experiment, we employed a noise measurement scheme[16] (see Fig. 2a), which utilized a small cantilever with a sharp (50 nm radius) conductive tip, mounted in a temperature controlled can within a vacuum chamber. The cantilever is driven at its resonance frequency (~160 kHz), via a piezoelectric and a self-driven oscillator circuit. Samples were 0.5 μm thick PVAc films (glass transition: $T_g$ = 308 K), which were spin-coated onto a metal substrate and annealed[14, 15]. When a voltage bias is applied between tip and substrate, fluctuations in the sample polarization beneath the tip produce proportional variations in the cantilever resonance frequency[16] which are detected using frequency modulation techniques[17]. For small tip heights, z, above the sample surface, the resonance frequency is most sensitive to a small region directly beneath the tip. For z = 30 nm and bias = 4V, we numerically calculated the maximum electric field: $E_{max} = 8.5 \times 10^6$ V/m, and the effective probed volume: $\Omega \sim 2 \times 10^{-17}$ cm$^3$, and depth: 40 nm. The latter is larger than the length scales below which surface effects dominate the dynamics[18].

With tip height fixed using feedback to reset the resonance frequency between each set of data points (every 16-65 s), time-series of PVAc polarization fluctuations were recorded at various temperatures slightly below $T_g$ and Fourier analyzed to obtain a power spectrum. Spectra averaged over long periods (>1 day) showed the expected power-law[15], $S(f) \sim f^{-\gamma}$. However, anomalous transient deviations were observed. An example is shown in figure 2b. Successive spectra, each measured over a one hour period show significant changes in shape. Most striking, is the growth and subsequent reduction of a spectral feature, which is similar in shape to a slightly broadened Lorentzian (also shown). Similar spectral features, which persisted for up to a few hours, were seen many times in the temperature range 298-302 K, and frequency range 0.05-5Hz in several samples. These spectral features provide convincing evidence for heterogeneity, which is however, transient. Analysis of the average lifetime of these heterogeneities was carried out by studying the time-autocorrelation function of the local spectral exponent $(-d\ln(S)/d\ln(f))$[15]. As shown in figure 3, heterogeneity lifetime was comparable at each temperature to the measured[14] average ($\alpha$) relaxation time. This result demonstrates a direct link of these (~ 3-4 decades) faster processes to the primary relaxation, and matches the behavior of heterogeneities found at higher temperatures and closer to the $\alpha$–peak frequency[8]. We note that the $\beta$ peak is 100 times smaller and 10 decades higher in frequency[19] and therefore unlikely to be relevant to these measurements.

Associated with the appearance of pronounced spectral features, we frequently observed random telegraph switching (RTS) between two to four discrete levels in the polarization time-series. See figure 4. Like the spectral features, the RTS were in all cases transient, persisting for times ranging from seconds (few cycles) to at most a few hours (thousands of cycles), and in some cases appearing and disappearing several times. The average switching rates of different observed RTS varied over the observable range (factor of ~ 100), clearly demonstrating dynamical heterogeneity. Often, as seen in Fig. 4a, RTS activity or rates were modulated in time, sometimes by discrete transitions, possibly due to other RTS.

Recent NMR studies[20] suggest that orientational relaxation in PVAc occurs via a random walk of ~ 10° angular jumps, with a typical jump time shorter than the $\alpha$ relaxation time. The RTS such as those in fig. 4b, with at least 4 repeatedly visited levels, can be modeled by a cluster whose orientation makes moderate sized jumps in a multiple-well orientational potential. In support of this picture, we note that changes in the applied electric field produced small but reproducible changes in the RTS level occupancy probabilities, consistent with changes in cluster dipole-moment or dipole-moment orientation (EVR and NEI to be published). These were used to infer cluster dipole moments which were found to be several times the dipole moment of a single PVAc monomer and were compatible with recent PVAc experiments[6] and theory[21] which found cluster sizes of 30-90 monomers.

In order to analyze the RTS kinetics, distributions of time durations were extracted from individual long-lived RTS series with two dominant levels. For a thermally activated two-state process, the time durations, t, spent in each configuration would be expected to be exponentially distributed, e.g. $N(t)=N_0 \exp(-t/\tau)$. However, for all RTS studied (see figure 5), the distributions could be fit well with a stretched-exponential, $N(t) = N_0 \exp(-t/\tau)^\beta$, with $0.4 < \beta < 0.6$, similar to the values found for bulk dielectric relaxation[14]. The $\tau$ values varied widely (factor of 100) for different RTS, again demonstrating heterogeneity. One possibility is that this stretching arises from discrete modulation of the rates (see Fig. 4a) of an underlying exponential process. To test this idea, long RTS time series were sliced into shorter sections (e.g. 200 s) corresponding to observation times which are short compared to the $\alpha$-relaxation time. For some RTS, distributions appeared exponential in selected shorter sections. As shown in the Fig. 5 inset, with increasing observation time the average value of $\beta$ decreased for these RTS, indicating an evolution with time toward a more stretched response. Other RTS remain highly stretched even for 200 s observation times, consistent with a more rapid modulation of rates. A simulated, intrinsically stretched-exponential, RTS showed time-independent $\beta$ (as shown) and no exponential short time behavior, indicating these effects were statistically significant.

In summary, using a nanoscale probe of dielectric fluctuations in PVAc, we have observed spontaneous switching between discrete polarization levels near the glass transition. These results directly demonstrate that the idea of cooperativity is applicable to molecular glasses. Cooperativity takes the form of transient molecular clusters, a few nanometers in size, which reorient (up to thousands of times) between 2 to 4 (or more) distinct configurations, and may be modeled by thermal-activation in a multiple-well orientational potential. Clusters had a wide range of characteristic switching rates, direct evidence that the broadening of the α–relaxation process arises from heterogeneity. The lifetime of heterogeneities was comparable to the α–relaxation time. Individual cluster configurations exhibited stretched-exponential kinetics with some showing a tendency toward exponential kinetics for observation times short compared with the α–relaxation time. This behavior was linked with the observed discrete modulations of switching rates, which might be understood in terms of a slowly fluctuating environment[22].

This research was supported by NSF Division of Materials Research, and the Petroleum Research Fund administered by the American Chemical Society. We thank M. D. Ediger for helpful discussions and K. Sinnathamby for assistance.

**Correspondence and requests for materials should be addressed to N. E. Israeloff (email: Israeloff@neu.edu).**

Figure 1. Heterogeneous scenario. **a,** The imaginary component of the dielectric susceptibility, $\varepsilon''$, is assumed to arise from a superposition of Debye peaks. In a mesoscopic sample (**b**), the individual Debye peaks become apparent as spectral features, particularly in the high-frequency tail of the $\alpha$–peak. In the thermal noise spectrum (**c**) the Debye peaks appear as Lorentzians, which also produce spectral features.

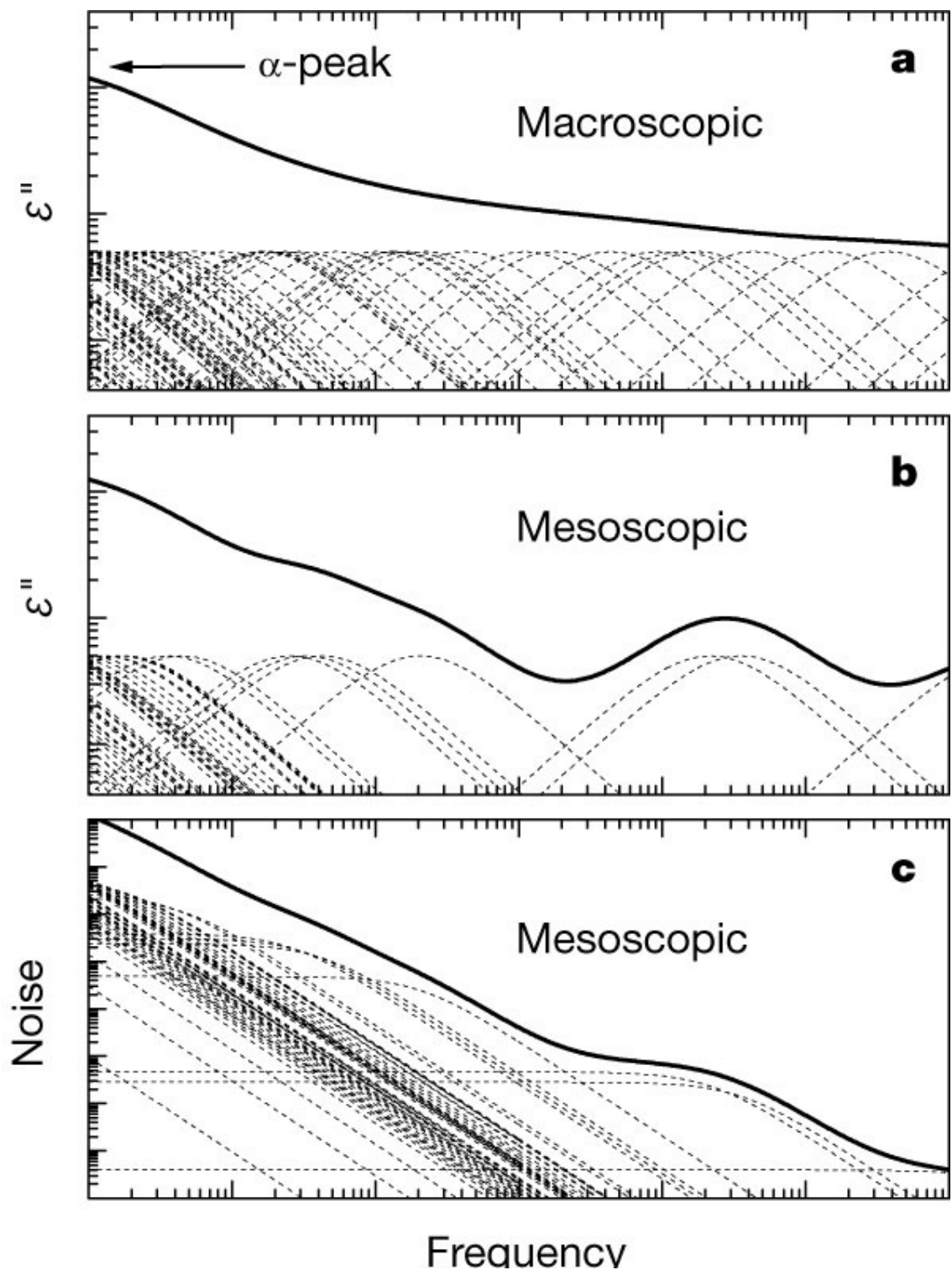

Figure 2. Noise spectra for PVAc at T=299 K. **a,** The AFM measurement setup. **b,** Successive polarization noise spectra (vs. frequency) show the appearance and subsequent disappearance of a spectral feature. Each spectrum is a 1-hour average with the relative starting time indicated. The dashed line shows a Lorentzian spectrum.

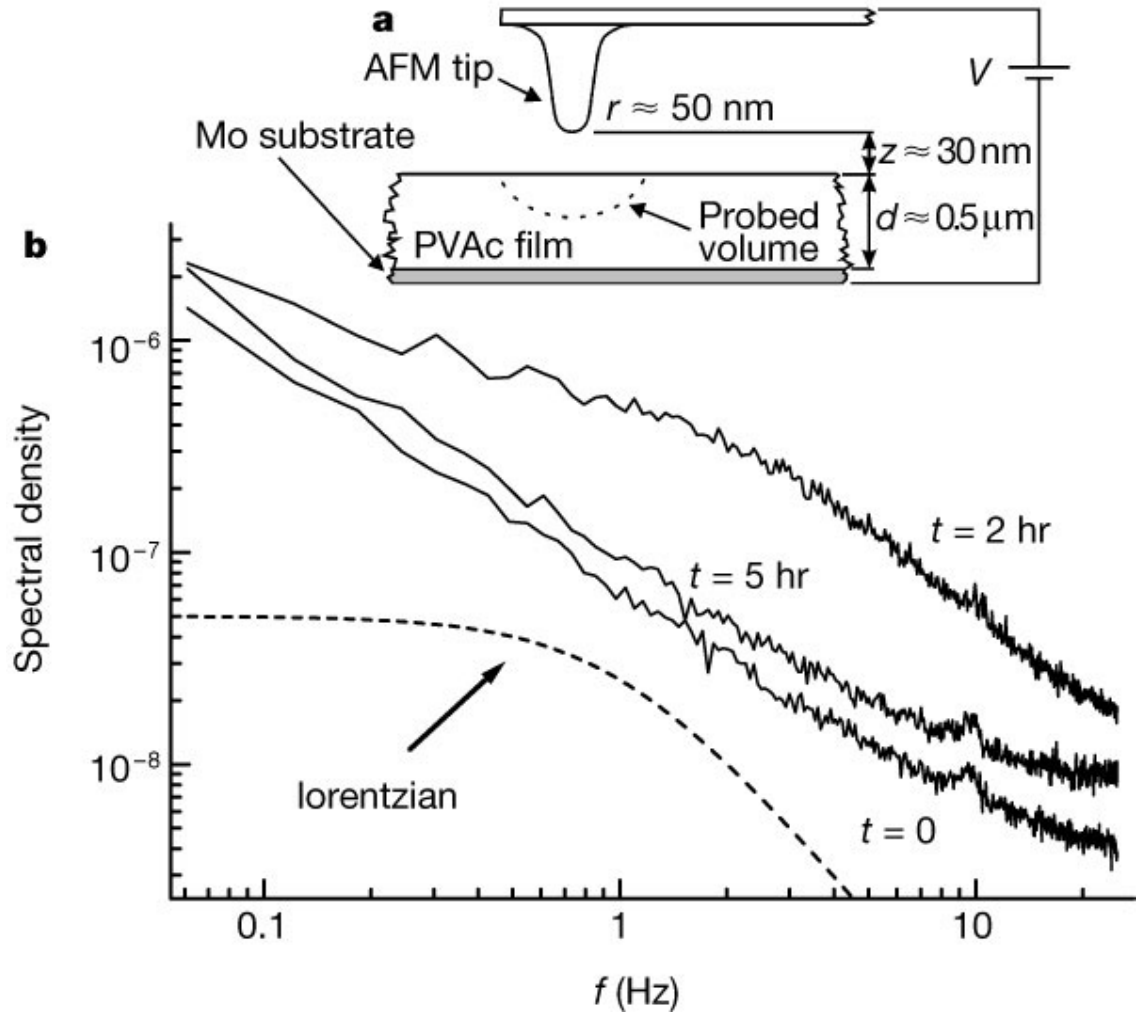

Figure 3. Lifetime of heterogeneities. The autocorrelation function of the local power spectral exponent[15] is shown for three temperatures with exponential fits and characteristic times. Inset: Heterogeneity lifetime and $\alpha$–relaxation time[14] are plotted vs. inverse temperature.

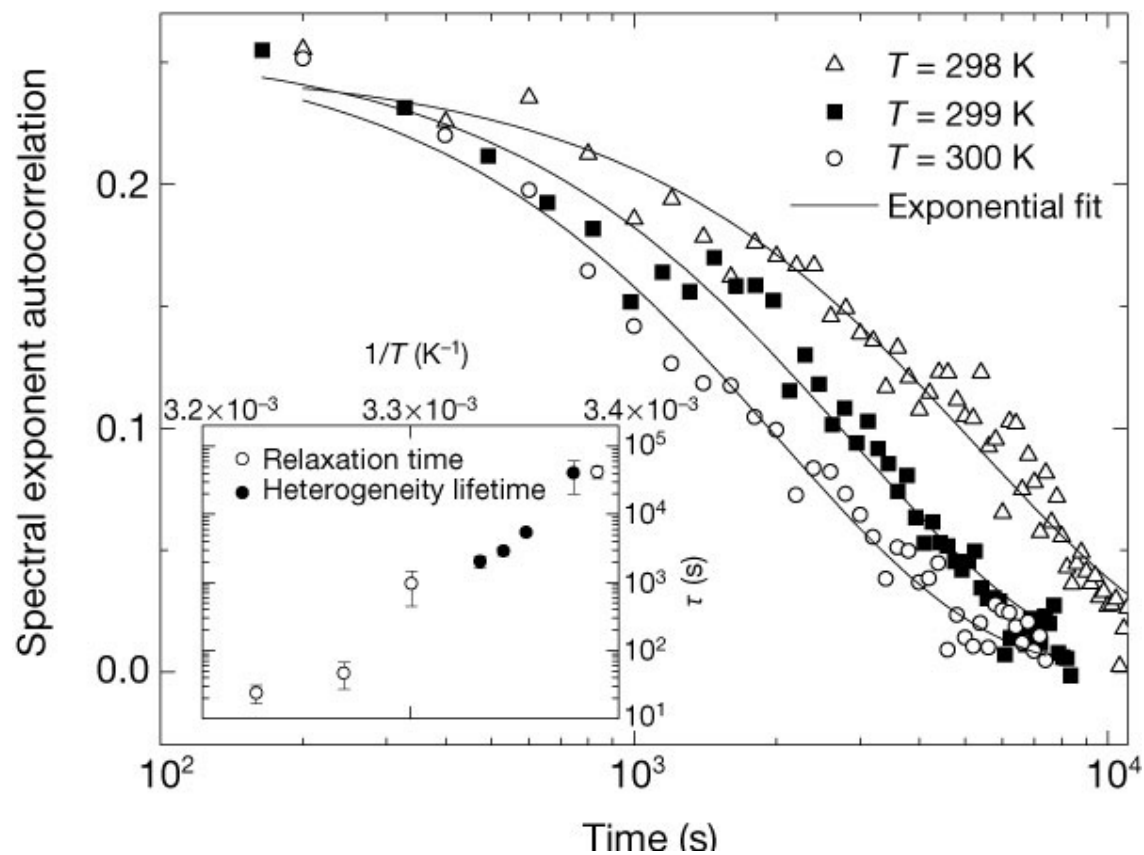

Figure 4. Time series of polarization. Two 50 s series showing random telegraph switching (RTS) noise. **a,** At 299 K, the rate of switching between two levels appears modulated. **b,** An example of a four-state switching at 300K is shown.

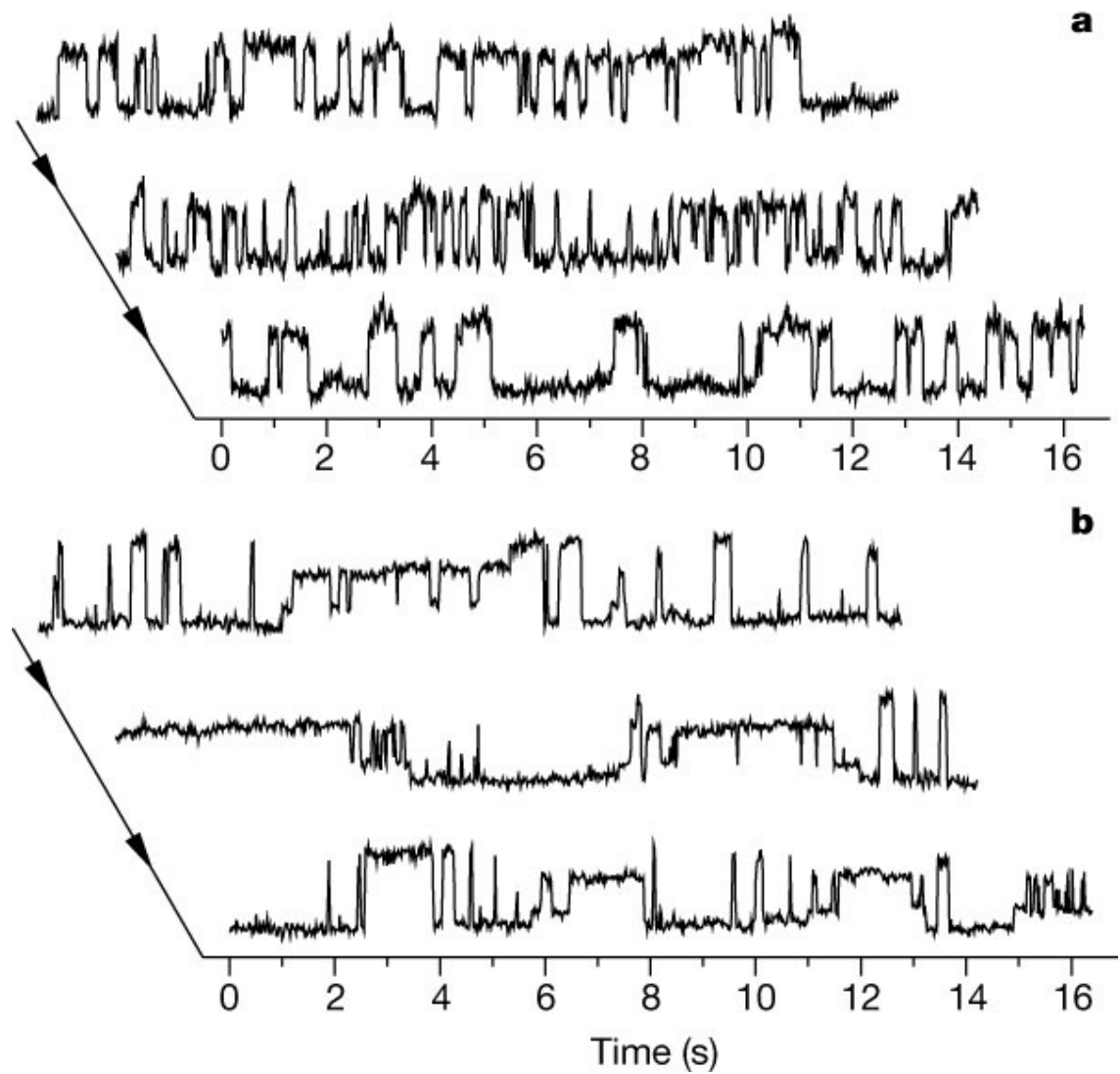

Figure 5. Individual cluster kinetics. Histograms of event durations for upper and lower states of a long-lived two-state RTS observed for 3200 s at 299 K are shown (points) with stretched exponential fit (solid line). An exponential would appear as a straight-line on this plot. The fit parameters are indicated. Inset: Average stretching exponent vs. observation time for the up-state data and a simulated intrinsic stretched-exponential RTS.

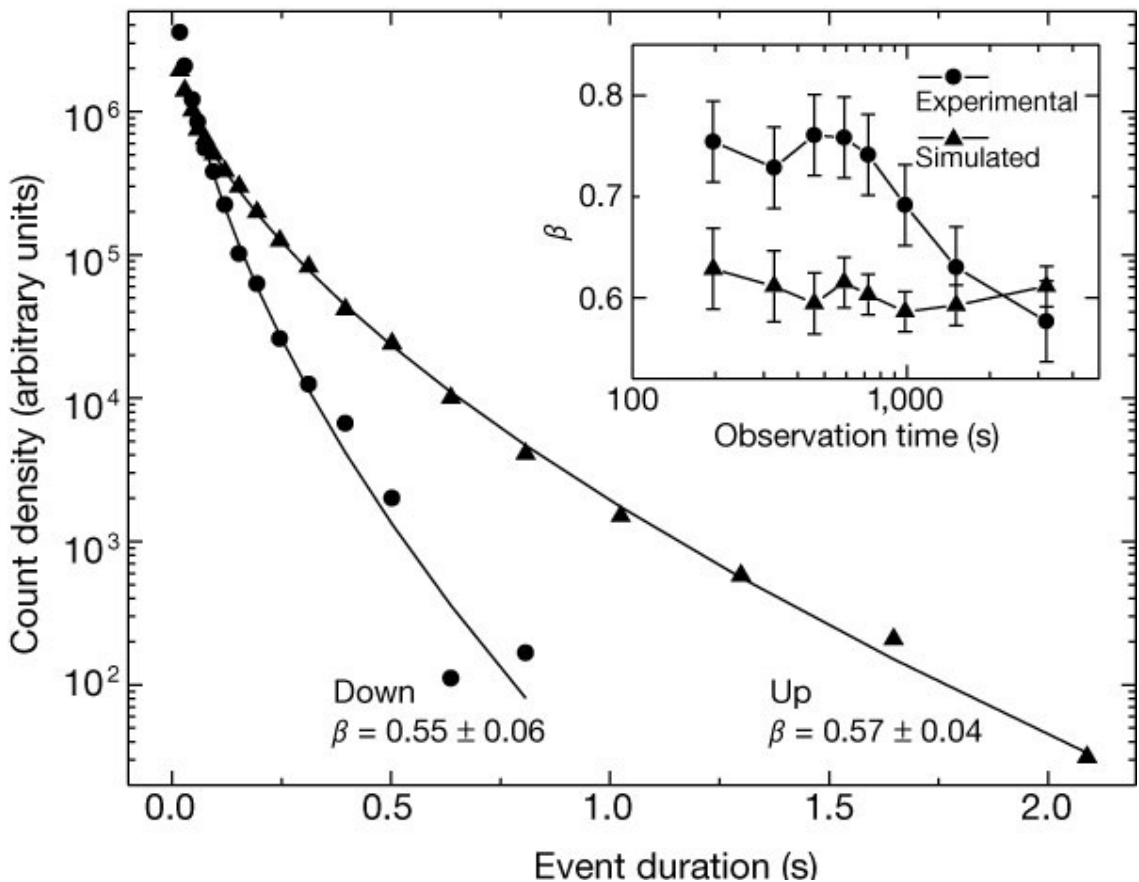